\documentclass[11pt]{article}
\newcommand{\Comment}[1]{{}}
\usepackage{amsfonts,amsthm,amsmath,amssymb,slashed}
\usepackage[textwidth = 430 pt, textheight = 630 pt]{geometry}
\usepackage{color}

\Comment{\usepackage{color}
\definecolor{MyDarkBlue}{rgb}{0.15,0.15,0.45}
\usepackage[linktocpage=true]{hyperref}
\hypersetup{
colorlinks=true,
citecolor=MyDarkBlue,
linkcolor=MyDarkBlue,
urlcolor=MyDarkBlue,
pdfauthor={Renato Costa and Horatiu Nastase},
pdftitle={Conformal inflation from the Higgs},
pdfsubject={hep-th}
}

\usepackage[numbers,sort&compress]{natbib}
\usepackage{hypernat}}
\usepackage{graphicx}
\usepackage{cite}

\newcommand\ignore[1]{}
\def\one{{\,\hbox{1\kern-.8mm l}}}

\def\a{\alpha}\def\b{\beta}

\def\d{\partial}

\newcommand{\Cset}{{\,\,{{{^{_{\pmb{\mid}}}}\kern-.45em{\mathrm C}}}}}

\newcommand{\be}{\begin{equation}}
\newcommand{\bea}{\begin{eqnarray}}

\newcommand{\ee}{\end{equation}}
\newcommand{\eea}{\end{eqnarray}}

\begin{document}

\renewcommand{\thefootnote}{\fnsymbol{footnote}}

\makeatletter
\@addtoreset{equation}{section}
\makeatother
\renewcommand{\theequation}{\thesection.\arabic{equation}}

\rightline{}
\rightline{}


\vspace{10pt}


\begin{center}
{\LARGE \bf{\sc Conformal inflation from the Higgs}}
\end{center} 
 \vspace{1truecm}
\thispagestyle{empty} \centerline{
{\large \bf {\sc Renato Costa}}\footnote{E-mail address: \Comment{\href{mailto:renato.costa87@gmail.com}}{\tt renato.costa87@gmail.com}}
{\bf{\sc and}}
{\large \bf {\sc Horatiu Nastase}}\footnote{E-mail address: \Comment{\href{mailto:nastase@ift.unesp.br}}{\tt nastase@ift.unesp.br}}
                                                  }

\vspace{.5cm}


\centerline{{\it 
Instituto de F\'{i}sica Te\'{o}rica, UNESP-Universidade Estadual Paulista}} \centerline{{\it 
R. Dr. Bento T. Ferraz 271, Bl. II, Sao Paulo 01140-070, SP, Brazil}}

\vspace{1truecm}

\thispagestyle{empty}

\centerline{\sc Abstract}

\vspace{.4truecm}

\begin{center}
\begin{minipage}[c]{380pt}
{\noindent We modify the conformal inflation set-up of \cite{Kallosh:2013hoa}, with a local Weyl invariance, a dynamical Planck scale 
and an $SO(1,1)$ invariance at 
high energies, in order to be able to identify the physical scalar with the physical Higgs at low energies, similar to the cyclic Higgs models 
of \cite{Bars:2013vba}. We obtain a general class of exponentially-corrected potentials that gives a generalized Starobinsky model with an infinite series 
of $R^p$ corrections, tracing a line in the $(r,n_s)$ plane for CMB fluctuations, with an $r$ that can be made to agree with the recent BICEP2 measurement. 
Introducing a coupling different from the conformal value, thus breaking the local Weyl symmetry, leads nevertheless
to a very strong attractor motion towards the generalized Starobinsky line.
}
\end{minipage}
\end{center}

\vspace{.5cm}

\setcounter{page}{0}
\setcounter{tocdepth}{2}

\newpage

\renewcommand{\thefootnote}{\arabic{footnote}}
\setcounter{footnote}{0}

\linespread{1.1}
\parskip 4pt



\section{Introduction}
\ \ \ \ \

With the latest results from the Planck experiment \cite{Ade:2013uln} and the recent BICEP2 measurement of $r$ \cite{Ade:2014xna}, 
cosmology and in particular inflation has finally started to become a testable field, 
since now we can rule out more and more models. Planck showed a preference towards inflation models with a plateau, as opposed to $\lambda_n\phi^n$
potentials, and in particular the Starobinsky model is close to the center of its allowed region in the $(r,n_s)$ plane. But the BICEP2 experiment 
favors instead large values of $r$, in the 0.15-0.25 range, seeming to exclude the simple Starobinsky model, with typically $r\sim 0.01$. 
The Starobinsky model is 
understood as inflation due to the presence of an $R^2$ term in the action, but it is equivalent to a usual Einstein plus scalar model with a potential that is an 
exponentially-corrected plateau. We will review this mechanism and generalize it to more complicated $f(R)$ actions, in the next section, in particular 
obtaining a generalization that allows for large $r$ compatible with BICEP2, tracing a line in the $(r,n_s)$ plane. 

But given that we want a plateau at large scalar field displacement values, where inflation happens, it is relevant to ask whether there is a symmetry
at high energies that forces the model to have a plateau, in which case inflation would arise naturally, avoiding some non-naturalness arguments raised by 
\cite{Ijjas:2013vea} (see \cite{Guth:2013sya} for counter-arguments to those criticisms). 
If we are looking for such a symmetry, one possible answer that comes to mind is local Weyl symmetry (symmetry under rescaling the metric and 
the scalar fields by a local conformal factor), which in the Standard Model with masses given by VEVs of the Higgs seems to be broken at the fundamental
level (not considering VEVs) only by the Planck mass $M_{\rm Pl}$. That can be remedied however by considering that $M_{\rm Pl}$ is also an effective coupling 
coming from the VEVs of scalars, leading to a theory with no {\em a priori} mass scales. To avoid issues with the experiments
on the time variation of the Planck scale, one is led to models with 2 scalar fields and the local Weyl symmetry, as in 
\cite{Kallosh:2013hoa,Kallosh:2013maa} (following earlier work by \cite{Ferrara:2010in,Bars:2011mh,Bars:2011aa,Bars:2012mt,Kallosh:2013lkr,Kallosh:2013pby};
note that for \cite{Ferrara:2010in,Kallosh:2013lkr,Kallosh:2013pby} the initial motivation was superconformal symmetry, whereas for 
\cite{Bars:2011mh,Bars:2011aa,Bars:2012mt} the initial motivation was from 2-time physics models. The two are however related since for both we have an 
$SO(4,2)$ symmetry, conformal in 3+1 dimensions vs. Lorentz in 4+2). 
If we impose also an $SO(1,1)$ symmetry (can be thought of as a remnant of breaking of $SO(4,2)\rightarrow SO(3,1)\times SO(1,1)$) 
between the 2 scalars at large field value
(equivalent of high energies in terms of the evolution of the Universe), and simply deform away from it, 
we are led to a theory with a plateau, giving naturally inflation. Moreover, one obtains the Starobinsky model predictions, basically because the 
scalars are related to the canonical scalar in an exponential way $\phi_i\sim e^{c\varphi}$ at large $\varphi$, and the $SO(1,1)$ symmetry cancels the 
leading contribution in the potential, as we will argue in detail in the current paper. The local Weyl invariance allows us to remove one of the scalars
by fixing a gauge, and be left with a single physical scalar. Naturally one obtains the original Starobinsky model, so it is important to look for 
generalizations that allow us to go to the Starobinsky line described above.

On the other hand, in \cite{Bars:2013yba,Bars:2013vba}, a similar model was considered, again with local Weyl invariance and two scalars, and the Planck 
mass coming from the same scalar VEVs, but without the $SO(1,1)$ symmetry between the scalars (except in the kinetic term, where it is essential!), 
where the motivation was different. We know of only one observed scalar so far, the recently discovered Higgs boson \cite{Aad:2012tfa,Chatrchyan:2012ufa}, 
so the natural question 
to be asked is whether we can identify the physical scalar with the Higgs. Indeed one can, and it was shown that if we want to keep only some 
terms natural for the Higgs in the potential, one obtains a cyclic cosmology. Basically, the reason is the same exponential relation to the 
canonical scalar $\phi_i\sim e^{c\varphi}$, but now because the lack of $SO(1,1)$ symmetry avoids the cancellation of the leading term and leaves us 
with a positive exponential potential $V\sim e^{a\varphi}$, which is one of the natural class of potentials for cyclic cosmology. Thus paradoxically, 
a similar set-up with or without $SO(1,1)$ at large energies leads to natural inflation or natural cyclic cosmology. 

In this paper we mix the two approaches by taking the best out of each one, noticing loopholes left by the two constructions, as well as 
generalizing the Starobinsky model by completing it to an infinite series summing to an $f(R)$, giving a line in the $(n_s,r)$ plane for CMB fluctuations. 
We note that allowing for $SO(1,1)$ at large field value and deforming away from it allows for a more general set-up than the one considered in 
\cite{Kallosh:2013hoa,Kallosh:2013maa}, in particular allowing us to naturally get the Higgs potential at low energies. Reversely, in 
\cite{Bars:2013yba,Bars:2013vba} one considered a quartic Higgs potential, but one can consider that the potential is modified at large field 
values to reach asymptotically the $SO(1,1)$ invariant form that can hold at high energies. We will see that in this case we obtain again a 
model giving predictions similar to the Starobinsky one in the $(r,n_s)$ plane, as in \cite{Kallosh:2013hoa,Kallosh:2013maa}, 
but now the model is  more general, and more importantly we can also obtain the generalized Starobinsky line, allowing us to agree with the 
$n_s$ value of Planck but also the $r$ value of BICEP2.     
Note that the idea of Higgs inflation is far from new, being in fact quite popular recently. Moreover, 
 there is in fact a model that can be thought of as a particular case of  the general set-up with 
a Higgs coupled to the Einstein action, otherwise with only a Higgs potential, the Bezrukov-Shaposhnikov model \cite{Bezrukov:2007ep}. It was shown in 
\cite{Bars:2013yba} that it admits a Weyl-symmetric formulation.

The paper is organized as follows. In section 2 we review the Starobinsky model, generalized to $f(R)$ actions, 
in particular showing that we can obtain an arbitrary value of $r$ through an $f(R)$ that goes over to $R^2$ in the inflating large $R$ region. 
Then we show that a class of exponentially-corrected potentials give rise to the same predictions (this observation was made in 
\cite{Ellis:2013nxa}, and in the context of supergravity models in \cite{Ferrara:2013rsa,Kallosh:2013yoa,Cecotti:2014ipa}; we learned about these references
after the paper was posted on the arXiv, from the referee; also \cite{Ellis:2014cma} made this prediction), also generalized to a line in the $(r,n_s)$ plane. 
In section 3 we present the class of models we will be considering, with local Weyl invariance, 
$SO(1,1)$ invariance and reducing to the Higgs potential at low energies. In section 4 we discuss various particular cases and the inflationary models 
arising from them. In section 5 we show that with a modification of the scalar-Einstein coupling $\xi$ away from the conformal value of $-1/6$ leads 
nevertheless to a strong attraction to the same Starobinsky line. In section 6 we conclude.

\section{Starobinsky models and a general class of potentials.}

The Starobinsky model is obtained by adding an $R^2$ correction term to the Einstein-Hilbert action, 
and it is equivalent to adding a scalar with a potential that is an exponentially corrected plateau. 
We will explain this mechanism by generalizing to an $R^p$ correction, where $p$ is an arbitrary positive real number. We start with the action
\be
S=\frac{M_{\rm Pl}^2}{2}\int d^4x\sqrt{-g}(R+\beta_{p+1}R^{p+1}).
\ee
It can be rewritten by introducing an auxiliary scalar field $\a$ as 
\be
S=\frac{M_{\rm Pl}^2}{2}\int d^4x\sqrt{-g}\left[R(1+(p+1)\beta_{p+1}\a)-p\beta_{p+1}\a^{\frac{p+1}{p}}\right].
\ee
Defining the Einstein frame metric
\be
g_{\mu\nu}^E=[1+(p+1)\beta_{p+1}\a]g_{\mu\nu}\equiv \Omega^{-2}g_{\mu\nu}\;,
\ee
and using the general formula for a Weyl rescaling in $d$ dimensions
\be
R [g_{\mu\nu}]=\Omega^{-2}\left[R[g^E_{\mu\nu}]-2(d-1)g^{\mu\nu}_E\nabla^E_\mu\nabla^E_\nu\ln \Omega-(d-2)(d-1)g^{\mu\nu}_E(\nabla_\mu^E\ln\Omega)
\nabla_\nu^E\ln\Omega\right]\;,
\ee
we obtain the Einstein plus scalar action
\be
S=\frac{M_{\rm Pl}^2}{2}\int d^4x\sqrt{-g_E}\left[R[g_E]-\frac{3(p+1)^2\beta_{p+1}^2}{2}g^{\mu\nu}_E\frac{\nabla_\mu^E\a\nabla_\nu^E\a}{[1+\beta_{p+1}(p+1)\a]^2}
-V(\a)\right]\;,
\ee
where 
\be
V(\a)=\frac{p\beta_{p+1}\a^{\frac{p+1}{p}}}{\left[1+(p+1)\b_{p+1}\a\right]^2}.
\ee
Note that the auxiliary scalar has become dynamical, having a kinetic term.
Defining the canonical scalar by 
\be
d\phi=\sqrt{\frac{3}{2}}\frac{(p+1)\b_{p+1} d\a}{1+\beta_{p+1}(p+1)\a}\Rightarrow 
\phi=\sqrt{\frac{3}{2}}\ln [1+\beta_{p+1}(p+1)\a]\;,\label{canscalar}
\ee
the potential is
\be
V(\phi)=p\b_{p+1}\left[\frac{e^{\sqrt{\frac{2}{3}}\frac{\phi}{M_{\rm Pl}}}-1}{\b_{p+1}(p+1)}\right]^{\frac{p+1}{p}}
e^{-2\sqrt{\frac{2}{3}}\frac{\phi}{M_{\rm Pl}}}.\label{starobpot}
\ee
The potential is real and positive definite for $\phi\geq 0$, but in order for it to continue to be also for
$\phi< 0$, we must have $p=1/(2k+1)$, with $k$ an integer. Then at large field value
$\phi\rightarrow\infty$, we obtain
\be
V\simeq \frac{p\b_{p+1}}{[(p+1)\b_{p+1}]^{\frac{p+1}{p}}}e^{\frac{1-p}{p}\sqrt{\frac{2}{3}}\frac{\phi}{M_{\rm Pl}}}
\left[1-\frac{p+1}{p}e^{-\sqrt{\frac{2}{3}}\frac{\phi}{M_{\rm Pl}}}+...\right]\;,
\ee
which means that we get an exponentially-corrected plateau only for $p=1$, otherwise there is a dominant overall exponential.

This method can be used for a general $f(R)$ action, but we will use it to reproduce a more general expansion at $\phi\rightarrow \infty$. 
We see that for the Starobinsky model $p=1, k=0$, as well as for the generalized Starobinsky model with general integer $k$, the power in the exponential 
correcting the plateau is $\sqrt{\frac{2}{3}}$. 

The above model suggests considering the more general expansion at $\phi\rightarrow \infty$,\footnote{This more general model and its 
predictions were considered also in \cite{Ellis:2013nxa} in a general setting. In \cite{Ferrara:2013rsa,Kallosh:2013yoa,Cecotti:2014ipa}, this model and its 
predictions were analyzed in the context of supergravity models, in particular arising from a certain K\"{a}hler potential. We thank the referee for pointing out
these references to us. See also \cite{Ellis:2014cma} for a recent treatment of the same model. Our interest in this general 
potential is in obtaining it from the model for conformal inflation from the Higgs, in the following sections.}
\be
V\simeq A\left[1-ce^{-a\frac{\phi}{M_{\rm Pl}}}\right].\label{generalpot}
\ee
For this model, the inflationary slow-roll parameters are 
\bea
\epsilon&\equiv& \frac{M_{\rm Pl}^2}{2}\left(\frac{V'(\phi)}{V(\phi)}\right)^2=\frac{a^2c^2}{2}e^{-2a\frac{\phi}{M_{\rm Pl}}}\ll |\eta|\cr
\eta&=&M_{\rm Pl}^2\frac{V''(\phi)}{V(\phi)}=-a^2ce^{-a\frac{\phi}{M_{\rm Pl}}}.
\eea
Then the parameters $(n_s,r)$ of the spectrum of CMB fluctuations are given by 
\bea
n_s-1&=&-6\epsilon+2\eta\simeq 2\eta=-2a^2ce^{-a\frac{\phi}{M_{\rm Pl}}}\cr
r&=&16\epsilon=8a^2c^2e^{-2a\frac{\phi}{M_{\rm Pl}}}=\frac{2}{a^2}(n_s-1)^2.
\eea
To fix the value of $e^{-a\frac{\phi}{M_{\rm Pl}}}$, we consider the number of e-foldings, $N_e$. Note that the value of $\phi$ is the value at horizon 
crossing (crossing outside the horizon, to return inside it now) for the scale of the CMB fluctuations. Inflation happens at large $\phi$, decreasing as 
the Universe expands, so the start of inflation can correspond to a $\phi_0$ which can be greater than $\phi$, but for simplicity we will consider them 
equal. Then 
\bea
N_e&=&-\int_{\phi_0/M_{\rm Pl}}^{\phi_f/M_{\rm Pl}}\frac{d\phi/M_{\rm Pl}}{\sqrt{2\epsilon}}=-\frac{1}{ac}\int_{\phi_0/M_{\rm Pl}}^{\phi_f/M_{\rm Pl}}
e^{a\frac{\phi}{M_{\rm Pl}}}\simeq \frac{1}{a^2c}e^{a\frac{\phi}{M_{\rm Pl}}}\cr
&=&\frac{1}{-\eta}=\frac{2}{1-n_s}\;,
\eea
where we have used the fact that $e^{a\phi_f/M_{\rm Pl}}\sim 1$ in order for inflation to end, so we have neglected this term in $N_e$. 

In conclusion, we obtain the constraints
\bea
1-n_s&=&\frac{2}{N_e}\cr
r&=&\frac{2}{a^2}(n_s-1)^2\cr
&=&\frac{8}{a^2N_e}.
\eea
Then, for instance with 50 e-folds, $N_e=50$, $n_s$ is fixed to be $n_s\simeq 0.9600$, which is in the middle of the allowed region by the Planck experiment 
\cite{Ade:2013uln}, while 60 e-folds gives $n_s\simeq 0.9667$, again compatible with the data (for instance, Planck + WMAP gives $0.9603\pm 0.0073$).
For concreteness, we will take $N_e=50$ from now on. With $a=\sqrt{2/3}$ as for the Starobinsky model (even in the case generalized to $R^{p+1}$), 
$r=3(n_s-1)^2$, so $n_s-1\simeq-1/25$ gives $r\simeq 0.005$, too small for the measurement of BICEP2 (which excludes such a virtually zero $r$ at at least $5\sigma$).

But within the context of this more general model, we can fit even the central value of $r$ of the BICEP2 experiment, of $r=0.20$ at $n_s-1\simeq
-1/25$, by $r\simeq 2/(25a)^2\sim 0.20$ for $a\sim 1/8$.

To complete the analysis of (\ref{generalpot}), we consider the running of $n_s$, 
\be
\frac{dn_s}{d\ln k}=-16\epsilon\eta+24\epsilon^2+2\xi^2\;,
\ee
where
\be
\xi^2\equiv M_{\rm Pl}^4\frac{V' V'''}{V^2}.
\ee
Then for (\ref{generalpot}) we have
\be
\xi^2\simeq c^2a^4e^{-2a\frac{\phi}{M_{\rm Pl}}}\;,
\ee
and $\xi^2\sim {\cal O}(e^{-2a\frac{\phi}{M_{\rm Pl}}})\gg \epsilon\eta, \epsilon^2$, so
\be
\frac{dn_s}{d\ln k}\simeq 2\xi^2\simeq 2c^2a^4e^{-2a\frac{\phi}{M_{\rm Pl}}}\simeq \frac{(n_s-1)^2}{2}.
\ee
Finally, the CMB normalization gives \cite{wmap9}
\be
\frac{H^2_{\rm inf}}{8\pi^2\epsilon M_{\rm Pl}^2}\simeq 2.4\times 10^{-9}\;,
\ee
where $H^2_{\rm inf}=V_{\rm inf}/3M_{\rm Pl}^2$. In our case, this is $A/3M_{\rm Pl}^2$, so that the constraint is 
\be
\frac{A/M_{\rm Pl}^4}{24\pi^2\epsilon}\simeq 2.4\times 10^{-9}.
\ee
Replacing $\epsilon$ by $r/16$, we get 
\be
\frac{A}{M_{\rm Pl}^4}\simeq \frac{3\pi^2}{2}r\times 2.4\times 10^{-9}.
\ee
With the central value of $r\simeq 0.2$ of BICEP2, we get 
\be
\frac{A}{M_{\rm Pl}^4}\simeq 7.1\times 10^{-9}\Rightarrow A^{1/4}\simeq 10^{-2}M_{\rm Pl}.
\ee
For the potential of the Starobinsky model in (\ref{starobpot}) with $p=1$, this gives the somewhat unnatural
\be
\beta_{2}\approx 1.4\times 10^8 M_{\rm Pl}^{-2}\simeq(10^{-4}M_{\rm Pl})^{-2}.
\ee

We now return to the question of how to obtain an $f(R)$ model, generalizing the Starobinsky model, that still allows for an $a$ as small as $\sim 1/8$. 
It was essential that we had the canonical scalar related to the scalar $\a$ by (\ref{canscalar}) in order to obtain $a=\sqrt{\frac{2}{3}}$.

Consider $M_{\rm Pl}=1$ for the moment, in order not to clutter formulas. We want to generalize (\ref{canscalar}) to 
\be
1+\a=e^{\sqrt{\frac{2}{3}}\frac{\phi}{b}}\Rightarrow d\phi=\sqrt{\frac{3}{2}}b\frac{d\a}{1+\a}\;,
\ee
which gives $a=\sqrt{\frac{2}{3}}\frac{1}{b}$. We see that then the kinetic term must come in the Jordan frame from a factor 
$R[1+\a]^b$. Since moreover we want the potential to be ($\b$ is a constant)
\be
V=\b\frac{(\a)^{2b}}{[1+\a]^{2b}}=\b\left[1-e^{-\sqrt{\frac{2}{3}}\frac{\phi}{b}}\right]^{2b}
\simeq \b\left[1-2be^{-\sqrt{\frac{2}{3}}\frac{\phi}{b}}\right]\;,
\ee
we obtain the action in Jordan frame 
\be
S=\frac{1}{2}\int d^4x\sqrt{-g}\left[R(1+\a)^b-\b\a^{2b}\right]\;,
\ee
and we want $(-)^{2b}$ to be real and positive, so we need $b=k+1$ to be an integer. 

To eliminate the auxiliary scalar $\a$, we solve for its equation of motion, obtaining
\be
R=2\b\frac{\a^{2b-1}}{(1+\a)^{b-1}}\;,
\ee
which we cannot invert for general $b$, but we see that by replacing $\a$ in the action we get an action of the type $f(R)$,
\be 
S=\frac{1}{2}\int d^4x \sqrt{-g}f(R).
\ee
At small curvature $R\rightarrow 0$, $\a\rightarrow 0$ (which means $\phi\rightarrow 0$ for the physical scalar, i.e. the end of inflation 
region $V\simeq 0$), $\a=(R/2\b)^{\frac{1}{2b-1}}$, giving
\be
f(R)\simeq R\left[1+\left(b-\frac{1}{2}\right)(2\b)^{-\frac{1}{2b-1}} R^{\frac{1}{2b-1}}+...\right]\;,
\ee
so we see that with $b=k+1$, we obtain the same first term in the expansion as in the generalized Starobinsky model above, but now the 
potential has a true plateau. At large curvature $R\rightarrow \infty$, $\a\rightarrow \infty$ (which means large physical scalar $\phi$, i.e. the 
inflating region), $\a=(R/2\b)^{\frac{1}{b}}$, giving
\be
f(R)\simeq \frac{R^2}{4\b}\;,
\ee
as in the Starobinsky model.

In conclusion, just by allowing for a completion of the Starobinsky model via a particular infinite series that sums to the above $f(R)$, we can 
obtain any value for $a$, thus any value for $r$ needed.

\section{Class of models with Weyl invariance and approximate $SO(1,1)$ invariance, reducing to Higgs.}

We start with a model with both local Weyl symmetry and $SO(1,1)$ invariance, where the Planck scale appears when choosing a gauge, and otherwise there is 
no dimensionful parameter,
\be
S=\frac{1}{2}\int d^4x\sqrt{-g}\left[\d_\mu\chi\d^\mu\chi-\d_\mu\phi\d^\mu\phi+\frac{\chi^2-\phi^2}{6}R-\frac{\lambda}{18}(\phi^2-\chi^2)^2\right].
\ee
The coupling of $1/6$ of the scalar to the Einstein term was chosen such as to have the conformal value, and the potential was chosen to be quartic, 
all in all giving the local Weyl symmetry under
\be
g_{\mu\nu}\rightarrow e^{-2\sigma(x)}g_{\mu\nu};\;\;\;
\chi\rightarrow e^{\sigma(x)}\chi, \;\;\;\;\;
\phi\rightarrow e^{\sigma(x)}\phi.
\ee
Since $\sqrt{-g}\rightarrow e^{-4\sigma(x)}\sqrt{-g}$, the potential needs to have scaling dimension 4 in the scalar fields. Moreover, imposing 
the $SO(1,1)$ symmetry of $\chi^2-\phi^2$, which is a Lorentz type symmetry (in fact, in \cite{Bars:2011mh,Bars:2011aa,Bars:2012mt} the motivation for 
this model was 2-time physics, written covariantly in $(4,2)$ Minkowski dimensions, and the $SO(1,1)$ is a remnant of the $SO(4,2)$ Lorentz invariance), 
we are forced to take also the potential $(\phi^2-\chi^2)$. The $SO(1,1)$ can also be obtained from a model with conformal invariance, again $SO(4,2)$, 
the initial motivation of \cite{Ferrara:2010in,Kallosh:2013lkr,Kallosh:2013pby}.

It would seem like $\chi$ is a ghost, but because we have a {\em local} Weyl symmetry, we can put one scalar degree of freedom to zero, such as to get rid
of the ghost. Choosing a gauge for the local scaling invariance will also necessarily introduce a scale, the Planck scale (it is obvious that in order to fix 
the scaling transformation we must choose a scale). Of course, since we have only one scale, physics is independent of this scale as it should be, since 
the scale is a gauge choice (calling $M_{\rm Pl}$ to be $1m$ or $10^{19}GeV$ or something else is meaningless unless there is an independent definition of 
what is $1m$ or $1GeV$ or some other scale).

One simple gauge choice is the {\em Einstein gauge} (E-gauge) 
$\chi^2-\phi^2=6M_{\rm Pl}^2$, which is solved in terms of the canonically normalized field $\varphi$ by
\be
\chi=\sqrt{6}M_{\rm Pl}\cosh\frac{\varphi}{\sqrt{6}M_{\rm Pl}};\;\;\;
\phi=\sqrt{6}M_{\rm Pl}\sinh\frac{\varphi}{\sqrt{6}M_{\rm Pl}}.\label{chiphivarphi}
\ee
In terms of it, the action is the Einstein action, with a canonical scalar and a cosmological constant,
\be
S=\int d^4x\sqrt{-g}\left[\frac{M_{\rm Pl}^2}{2}R-\frac{1}{2}\d_\mu\varphi\d^\mu\varphi-\lambda M_{\rm Pl}^4\right].\label{einstein}
\ee
Another simple gauge choice is the {\em physical Jordan gauge or c-gauge}, $\chi(x)=\sqrt{6}M_{\rm Pl}$. In terms of it, the action is 
\be
S=\int d^4x\sqrt{-g}\left[\frac{M_{\rm Pl}^2}{2}R\left(1-\frac{\phi^2}{6M_{\rm Pl}^2}\right)-\frac{1}{2}\d_\mu\phi\d_\mu\phi-\frac{\lambda}{36}
(\phi^2-6M_{\rm Pl}^2)^2\right].\label{actionJordan}
\ee
We can go to the Einstein metric by $g_{\mu\nu}^E=(1-\phi^2/6M_{\rm Pl}^2)g_{\mu\nu}^J$ and then define the canonically normalized field $\varphi$ by 
$d\varphi/d\phi=(1-\phi^2/6M_{\rm Pl}^2)^{-1}$, and reobtain (\ref{einstein}). But one thing one can note now is that the potential for the field $\phi$
with the conformal coupling to $R$ looks like the Higgs potential, just with scalar VEV $v^2=6M_{\rm Pl}^2$, instead of the experimentally 
known $v\simeq 246 GeV$. 

Otherwise one could identify $\phi$ with a physical Higgs field, but the Higgs VEV of $\sqrt{6}M_{\rm Pl}$ was the result of the $SO(1,1)$ symmetry, 
and in the Einstein frame it led to an exactly flat potential (cosmological constant). If we completely discard the $SO(1,1)$ symmetry as in 
\cite{Bars:2013yba,Bars:2013vba} and insist instead on having the Higgs potential from low energies, $\lambda(H^\dagger H-v^2)^2$, appear in the physical
Jordan gauge, we are led to $\lambda(\phi^2-\omega^2\chi^2)^2$, generalized with the addition of a term $\lambda'\chi^4$ (in order to have the most 
general quartic potential for $\phi$ and $\chi$) that reduces to a cosmological constant. Note however that for the kinetic terms and $R$-coupling 
terms we must insist on the same $SO(1,1)$ symmetric form, if we want to remain with only the physical Higgs and the scalar VEV after gauge fixing, 
which seems contrived if we completely abandon the symmetry in the potential.

The attraction of this construction is that now in the Lagrangean for the Standard Model coupled to gravity there are no explicit mass scales, 
and the fundamental theory is local Weyl-invariant. 
Masses in the Standard Model come from coupling to the Higgs, and the Higgs VEV and Planck mass now come from choosing a gauge in a Weyl-invariant 
theory. 

However, the absence of the $SO(1,1)$ symmetry means that we are naturally led towards a cyclic cosmology. Indeed, in the Einstein gauge
for the local Weyl symmetry defined by (\ref{chiphivarphi}), the potential is now 
\be
V\sim \lambda M_{\rm Pl}^4\left(\sinh^2\frac{\varphi}{\sqrt{6}M_{\rm Pl}}-\omega^2\cosh^2\frac{\varphi}{\sqrt{6}M_{\rm Pl}}\right)^2-\lambda'M_{\rm Pl}^4
\cosh^4\frac{\varphi}{\sqrt{6}M_{\rm Pl}}\approx C M_{\rm Pl}^4\exp\frac{4\varphi}{\sqrt{6}M_{\rm Pl}}\;,
\ee
i.e. an exponential of the canonical scalar at large field value. This generically leads to cyclic cosmology, and we see that the origin of this is 
the non-cancellation of the leading exponentials. In the $SO(1,1)$ symmetric form, there was no $\lambda'\chi^4$ term and $\omega=1$ at high field values, 
leading to a cancellation of the leading exponentials in $\cosh^2-\sinh^2$, giving in fact a constant. 

Besides the fact that requiring $SO(1,1)$ symmetric kinetic terms but no remnant of the symmetry in the potential is unnatural, we also have the issue that
general cyclic cosmologies are incompatible with a large value of the tensor to scalar ratio of CMB fluctuations, $r$ \cite{Ijjas:2013vea,Lehners:2013cka}, 
and the BICEP2 experiment finds a value close to $r\sim 0.2$ \cite{Ade:2014xna}. So it would be ideal to find a model that gives inflation at high 
energies, and a natural way for that is to have an approximate $SO(1,1)$ at high energies, which we take to translate into large field values. 

So we ask what is the most general potential depending on $\phi$ and $\chi$ 
with the desired properties: local Weyl symmetry, to reduce at large field values (large energies) 
to the $SO(1,1)$ symmetric form, and at small field values (small energies) to the Higgs form. From just local Weyl symmetry, the most general 
form is $f(\phi/\chi)\phi^4$, since $\sqrt{-g}\phi^4$ and $\phi/\chi$ are both local Weyl invariant. But we want also to reduce to the $SO(1,1)$ 
invariant form $(\phi^2-\chi^2)^2$ at large field values, so we must have 
\be
V=\lambda\left[\tilde f(\phi/\chi)\phi^2-\tilde g(\phi/\chi)\chi^2\right]^2\;,\label{fgtilde}
\ee
or in another parametrization 
\be
V=\lambda f(\phi/\chi)\left[\phi^2-g(\phi/\chi)\chi^2\right]^2.
\ee
Since in the Einstein gauge $\phi/\chi=\tanh \varphi/\sqrt{6}M_{\rm Pl}$ which goes to 1 at large $\varphi$, we must require $g(1)=1$ for 
$SO(1,1)$ symmetry at large field values. In \cite{Kallosh:2013hoa,Kallosh:2013maa} it was considered only $g(x)\equiv 1$, hence the 
possible connection with the Higgs was not considered. 

Now the condition that we obtain the Higgs potential at low field values (low energies) is that $g(0)\simeq \omega^2$, where $\omega=246 GeV/\sqrt{6}M_{\rm Pl}$.
Moreover, the function $g(x)$ must give at small energies subleading corrections to the Higgs potential. For the simplest function, a polynomial plus a 
constant, we can take
\be
g(x)=\omega^2+(1-\omega^2)x^n\;,\label{gofx}
\ee
and we must impose $n>2$. Indeed, if $n=1$ (and $f(x)=1$) we get at small field values a potential $\propto [\phi^2-(\omega^2+(1-\omega^2)\phi/\chi)\chi^2]^2$ and 
in the Einstein gauge at small field the leading terms in the potential become
\be
V\simeq \lambda\left[(1-\omega^2)(\varphi^2-\sqrt{6}\varphi M_{\rm Pl})^2-6\omega^2 M_{\rm Pl}^2\right]^2\;,
\ee
so the linear term will dominate over the quadratic term in the square bracket. In the $n=2$ case we can check that actually the good $\phi^2$ term cancels
in the square bracket and the potential is simply $V\simeq  36\lambda\omega^4M_{\rm Pl}^4$, so again is not good. For $n>2$ instead, at small field the 
higher order corrections are suppressed by $M_{\rm Pl}$, i.e. we get approximately
\be
V\simeq \lambda \left[\varphi^2-6\omega^2M_{\rm Pl}^2-6M_{\rm Pl}^2\left(\frac{\varphi}{\sqrt{6}M_{\rm Pl}}\right)^n\right]^2\simeq 
\left[\varphi^2-6\omega^2M_{\rm Pl}^2\right]^2\;,
\ee
where we have used that $\omega^2\ll 1$, and we have dropped all terms coming from higher orders in the expansion of $\cosh(\varphi/\sqrt{6}M_{\rm Pl})$ and
$\sinh(\varphi/\sqrt{6}M_{\rm Pl})$. 

Finally now, we can have an overall function $f(\phi/\chi)$, but if we want to have the Higgs potential at low energies, we need again to 
restrict its form. We write it as $f(x)=1+F(x)$, and a simple possible form for $F(x)$ would be a polynomial, $F(x)=c_n x^n$, but by the same logic as above, 
we need to have $n>2$, if not we modify the Higgs potential at small field values.
In the next section we will study the resulting inflation from these models, and we will see other possible relevant examples for $F(x)$.

We have not addressed the issue of how can these models arise from a fundamental theory? The potentials described need to be only effective potentials, i.e. 
including quantum corrections. As we saw, the scale for the corrections to the Higgs potential is the Planck scale, so naturally these are quantum gravity
corrections, which could in principle come, for example, from string theory. The quantum theory is supposed to be valid at large field values, which is 
generically the case for scalars arising from string models. Moreover, the energy density at large field values is very large, which again implies that we 
are in the quantum gravity region. At these high energies, we have as usual the local Weyl symmetry, but also the $SO(1,1)$ symmetry which gets deformed, so it 
would be natural to assume that $\varphi\rightarrow \infty$ is a special point that has the full symmetry, but the symmetry is not protected away from it. 
However, the understanding of this is left for future work.

Finally, instead of the scalar $\phi$ we have the Higgs, which is a doublet, whereas $\chi$ would be a singlet under $SU(2)$, and then $\phi$ from the 
above discussion, the field relevant for inflation, would be its norm $\phi=\sqrt{H^\dagger H}$. The action is then 
\bea
S&=&\frac{1}{2}\int d^4x\sqrt{-g}\left[\d_\mu\chi\d^\mu\chi-\d_\mu H^\dagger\d^\mu H+\frac{\chi^2-H^\dagger H}{6}R\right.\cr
&&\left.-\frac{\lambda}{18}\left(1+F(\sqrt{H^\dagger H}/\chi)\right)
\left(H^\dagger H-g(\sqrt{H^\dagger H}/\chi)\chi^2\right)^2+{\cal L}'_{SM}\right]\;,
\eea
where ${\cal L}'_{SM}$ is the Standard Model Lagrangean other than the Higgs kinetic and potential term.

\section{Inflation in these models.}

We next turn to inflation in these models. We start with models with $F(x)=0$ ($f(x)=1$). 
For the simple $g(x)$ in (\ref{gofx}), we obtain for the potential at $\varphi\rightarrow\infty$, using that $\omega^2\ll 1$,
\be
V\simeq 9(n-2)^2\lambda M_{\rm Pl}^4\left[1-2n e^{-\frac{2\varphi}{\sqrt{6}M_{\rm Pl}}}\right]\;,\label{vinfty}
\ee
which means, in the parametrization of section 2, that $a=\sqrt{2/3}$ as in the Starobinsky model, and $c=2n$. That gives
\bea
\epsilon&\simeq& \frac{4n^2}{3}e^{-\frac{4\varphi}{\sqrt{6}M_{\rm Pl}}}\cr
\eta&\simeq &-\frac{4n}{3}e^{-\frac{2\varphi}{\sqrt{6}M_{\rm Pl}}}\;,
\eea
and then $n_s-1\simeq 2\eta$, $r\simeq 16\epsilon$, but more relevantly we get 
\bea
1-n_s&\simeq&\frac{2}{N_e}\cr
r&\simeq&3(n_s-1)^2\simeq\frac{12}{N_e}\;,
\eea
exactly as in the Starobinsky model. As we saw for the parametrization in section 2, we also have $dn_s/d\ln k\simeq (n_s-1)^2/2$ and $
A^{1/4}\simeq 10^{-2}M_{\rm Pl}$, which translates now into $\lambda^{1/4}\sqrt{3(n-2)}\sim 10^{-2}$.

A two-parameter generalization with 
\be
g(x)=\omega^2(1-x^m)+x^n\;,
\ee
reducing to the previous case for $m=n$, gives however still (\ref{vinfty}) at $\varphi\rightarrow\infty$, so obtains nothing new.

Another simple case with $F(x)=0$ and 
\be
g(x)=\omega^2+(1-\omega^2)\sin\left(\frac{\pi}{2}x^n\right)\;,
\ee
that still satisfies the condition to interpolate between the Higgs potential and the inflationary potential,
gives for the potential at $\varphi\rightarrow\infty$
\be
V\simeq 36\lambda\left(1-\frac{n^2\pi^2}{4}e^{-\frac{2\phi}{\sqrt{6}M_{\rm Pl}}}\right).
\ee
Then we have again $a=\sqrt{2/3}$ as in the Starobinsky model, but now $c=n^2\pi^2/4$. We get therefore again the same Starobinsky model predictions
$r\simeq 3(1-n_s)^2$, $1-n_s\simeq 2/N_e$ and $dn_s/d\ln k\simeq (1-n_s)^2/2$. Again $A^{1/4}\simeq 10^{-2}M_{\rm Pl}$ translates into 
$\sqrt{6}\lambda^{1/4}\sim 10^{-2}$.

Finally, consider an $F(x)=c_p x^p$ with $p>2$ and $g(x)$ in (\ref{gofx}). Then the potential at $\varphi\rightarrow\infty$ is 
\be
V\simeq \lambda M_{\rm Pl}^4 9(n-2)^2(1+c_p)\left[1-2\left(n+\frac{p}{1+c_p}\right)e^{-\frac{2\phi}{\sqrt{6}M_{\rm Pl}}} \right]\;,
\ee
so once again we obtain the predictions of the Starobinsky model. 

We see that the predictions of these models are robust and generically give the same as the Starobinsky model. The reason is that 
we have functions of $\phi/\chi=\tanh(\varphi/\sqrt{6}M_{\rm Pl})$ which is $\simeq 1-2e^{-\frac{2\varphi}{\sqrt{6}M_{\rm Pl}}}$ at 
$\varphi\rightarrow\infty$, so any well-defined Taylor expansion in $\phi/\chi$ would give the same $a=\sqrt{2/3}$. 

In order to obtain a different $a$ we need functions which have a somewhat singular behaviour at $\phi/\chi\rightarrow 1$. One obvious, yet 
somewhat unnatural example for the function $F(x)$ that would give a general $a$ and preserves the Higgs potential at small $\varphi$
would be 
\be
F(x)= c_p\left\{\tanh\left[a\sqrt{\frac{3}{2}}\tanh^{-1}(x)\right]\right\}^p\;,
\ee
that gives at $\varphi\rightarrow\infty$
\be
F\simeq  c_p\left[1-2pe^{-a\frac{\varphi}{M_{\rm Pl}}}\right].
\ee
Note however that in the context of supergravity models, the resulting potential for $\varphi$ can easily appear (see e.g. \cite{Kallosh:2013yoa}, eq. 5.1).  
A more plausible example made up of only logs and powers is 
\be
F(x)=c_{\gamma,p}\left[\ln\left(\frac{2}{1+x^p}\right)\right]^\gamma.\label{singularF}
\ee
At $\varphi\rightarrow\infty$ it gives
\be
F\simeq c_{\gamma,p}p^\gamma e^{-\sqrt{\frac{2}{3}}\frac{\gamma\phi}{M_{\rm Pl}}}\;,
\ee
which implies a general $a=\gamma\sqrt{2/3}$. Moreover, at $\varphi\rightarrow 0$, we get
\be
F\simeq c_{\gamma,p}(\ln 2)^\gamma\left[1-\frac{\gamma}{\ln 2}\left(\frac{\varphi}{\sqrt{6}M_{\rm Pl}}\right)^p\right]\;,
\ee
so with $p>2$ we don't perturb the Higgs potential, and we just rescale the coupling $\lambda\rightarrow \lambda[1+c_{\gamma,p}(\ln 2)^\gamma]$. 

The general conditions on $F(x)$ can be described as $F(1-x)\propto x^{\a_1}$  and $F(x)\simeq c_1+c_2 x^p$, $p>2$, for $x\rightarrow 0$.

We note that with these $F(x)$'s, any normal $g(x)$ like the ones given above would do the job, 
since at $\varphi$, we need experimentally $a\sim 1/8$, so the corrections 
coming from $F(x)$ would be leading with respect to the corrections coming from $g(x)$, which have $a=\sqrt{2/3}$. 

In conclusion, with not too singular functions $g(x)$ and $F(x)$ we obtain the predictions of the Starobinsky model, but with ones with a more singular 
behaviour at $x=1$ like for instance  (\ref{singularF}) we can obtain the generalized Starobinsky model of section 2, with arbitrary $a$, being able to 
fit the BICEP2 data with $a\sim 1/8$. 

Note that in this section  we have been interested in models appearing naturally in the conformal inflation from the Higgs, though from the point of 
view of general supergravity models, one can obtain the generalized potential in section 2 from a K\"{a}hler potential for a coset construction
\cite{Ferrara:2013rsa,Kallosh:2013yoa,Cecotti:2014ipa}.

\section{General coupling $\xi$ and attractors.}

The exact $SO(1,1)$ symmetric model in the physical Jordan gauge is (\ref{actionJordan}), and now consider the deformation of 
the potential with the parametrization in terms of $\tilde f, \tilde g$ in (\ref{fgtilde}). The action in physical Jordan gauge 
then becomes
\bea
S&=&\int d^4x\sqrt{-g}\left[\frac{M_{\rm Pl}^2}{2}R\left(1-\frac{\phi^2}{6M_{\rm Pl}^2}\right)-\frac{1}{2}\d_\mu\phi\d^\mu\phi-\right.\cr
&&\left.-\left(\tilde f\left(\frac{\phi}{\sqrt{6}M_{\rm Pl}}\right)\phi^2-\tilde g\left(\frac{\phi}{\sqrt{6}M_{\rm Pl}}\right)6M_{\rm Pl}^2\right)^2\right].
\eea
But we want to study the case of a general coupling $\xi<0$ of the scalars to gravity $\xi \phi^2R$, away from the conformal coupling $\xi=-1/6$. 
For consistency we replace $-\phi^2/(6M_{\rm Pl}^2)$ with $\xi\phi^2/M_{\rm Pl}^2$ everywhere, to obtain
\bea
S&=&\int d^4x\sqrt{-g}\left[\frac{M_{\rm Pl}^2}{2}R\left(1+\xi\frac{\phi^2}{M_{\rm Pl}^2}\right)-\frac{1}{2}\d_\mu\phi\d^\mu\phi-\right.\cr
&&\left.-36\left(\tilde f\left(\frac{\sqrt{|\xi|}\phi}{M_{\rm Pl}}\right)\xi\phi^2+\tilde g\left(\frac{\sqrt{|\xi|}\phi}{M_{\rm Pl}}\right)
M_{\rm Pl}^2\right)^2\right].
\eea
We can go to the Einstein frame $g_{\mu\nu}^E=(1+\xi\phi^2)g_{\mu\nu}$ and obtain 
\bea
S&=&\int d^4x\sqrt{-g_E}\left[\frac{M_{\rm Pl}^2}{2}R[g_E]-\frac{1}{2}\left[\frac{1+\xi\phi^2/M_{\rm Pl}^2
+6\xi^2\phi^2/M_{\rm Pl}^2}{(1+\xi\phi^2/M_{\rm Pl}^2)^2}\right]g^{\mu\nu}_E\nabla_\mu^E\phi
\nabla_\nu^E\phi-\right.\cr
&&-\left.36(1+\xi\phi^2/M_{\rm Pl}^2)^2
\left(\tilde f\left(\frac{\sqrt{|\xi|}\phi}{M_{\rm Pl}}\right)\xi\phi^2+\tilde g\left(\frac{\sqrt{|\xi|}\phi}{M_{\rm Pl}}\right)
M_{\rm Pl}^2\right)^2\right];\,
\eea
and define a canonical scalar field $\varphi$ by
\be
\frac{d\varphi}{d\phi}=\frac{\sqrt{1+\xi\phi^2/M_{\rm Pl}^2+6\xi^2\phi^2/M_{\rm Pl}^2}}{1+\xi\phi^2/M_{\rm Pl}^2}\;,\label{phivarphi}
\ee
to finally obtain the action
\bea
S&=&\int d^4x\sqrt{-g_E}\left[\frac{M_{\rm Pl}^2}{2}R[g_E]-\frac{1}{2}g^{\mu\nu}_E\nabla_\mu^E\varphi\nabla_\nu^E\varphi-\right.\cr
&&-\left.36\left(\tilde f\left(\frac{\sqrt{|\xi|}\phi(\varphi)}{M_{\rm Pl}}\right)\xi\phi^2+\tilde g\left(\frac{\sqrt{|\xi|}\phi(\varphi)}{M_{\rm Pl}}\right)
M_{\rm Pl}^2\right)^2\right].
\eea
The potential is a function of $\frac{\sqrt{|\xi|}\phi(\varphi)}{M_{\rm Pl}}$ as in \cite{Kallosh:2013hoa,Kallosh:2013maa}, 
but of a form restricted by our conditions. The analysis then follws in a similar way. If $\xi$ is not extremely small such as to be able to ignore the 
non-minimal coupling of the scalar to gravity, then inflation will occur at $\varphi\rightarrow\infty$, like in the $\xi=-1/6$ case, because of a plateau 
behaviour. From (\ref{phivarphi}), this is seen to be where $1+\xi\phi^2/M_{\rm Pl}^2\ll 1$, and in that region we obtain 
\be
\frac{d\varphi}{d\phi}\simeq\frac{\sqrt{6}|\xi|\phi/M_{\rm Pl}}{1+\xi\phi^2/M_{\rm Pl}^2}\;,\label{region}
\ee
solved by 
\be
\phi^2=\frac{M_{\rm Pl}^2}{|\xi|}\left(1-e^{-\sqrt{\frac{2}{3}}\frac{\varphi}{M_{\rm Pl}}}\right).\label{region2}
\ee
We see that this is again the same asymptotic functional form (except for the overall constant) as in the conformal case $\xi=-1/6$, which as we explained in 
the last section was the reason why we generically obtained the Starobinsky potential with $a=\sqrt{2/3}$, for functions $f(x)$ and $g(x)$
that are not too singular at $x=1$. 

Therefore by the same arguments as in the previous sections we will obtain the Starobinsky point also for this non-conformal $\xi<0$. The only remaining 
issue is to define the small value of $\xi$ at which we can consider chaotic inflation. The argument in \cite{Kallosh:2013hoa,Kallosh:2013maa} can be 
carried over with little modifications. For a smooth potential function at $\phi=0$ we can Taylor expand and thus consider only a quadratic potential 
$V\simeq m^2\phi^2/2$ and chaotic inflation for that, 
for which the number of e-folds is $N_e=\phi^2/4M_{\rm Pl}^2$, so for 60 e-folds we obtain $\phi_{60}\simeq 15M_{\rm Pl}$. 
Then the condition to be able to ignore the non-minimal coupling to gravity is if $|\xi|\phi_{60}^2/M_{\rm Pl}^2\ll 1$, so as to never need to reach the 
region (\ref{region}). That gives $|\xi|\ll 4\times 10^{-3}$. If however $|\xi|\gg 4\times 10^{-3}$, we are basically at the Starobinsky point. 
Therefore the Starobinsky point is a strong attractor in terms of a nonzero $\xi$, with even a small $\xi$ driving us away from the chaotic inflation 
point towards the Starobinsky point.

In the case of functions $f(x)$ and $g(x)$ that are sufficiently singular at $x=1$ to give a potential with a general $a$, the same analysis 
follows. For $|\xi|\ll 4\times 10^{-3}$ we can Taylor expand the potential at $\phi=0$ and obtain chaotic inflation, but for $|\xi|\gg 4 \times 10^{-3}$
we obtain (\ref{region2}). Combined with the condition $F(1-x)\propto x^{\a_1}$, we get in the inflation region
\be
F\left(\frac{|\xi|\phi^2}{M_{\rm Pl}^2}\right)\propto e^{-\sqrt{\frac{2}{3}}\frac{\a_1\varphi}{M_{\rm Pl}}}\;,
\ee
as before. So in the generalized Starobinsky case we also get a strong attractor behaviour towards the generalized Starobinsky line, exactly as in the case 
of the Starobinsky point.

\section{Conclusions}

In this paper we have studied the possibility to have the Higgs field as the inflaton in a model with local Weyl symmetry, where the Planck mass 
appears by fixing a gauge, and with $SO(1,1)$ invariance at large field values (in the inflationary region). We have shown that in these models, 
defined by two functions $f(x)$ and $g(x)$, generically
we obtain the predictions of the original Starobinsky model, but we can find functions that give a generalized version of the Starobinsky model, 
for which we can obtain any value of the tensor to scalar ratio of CMB fluctuations $r$, including the value of $\sim 0.2 $ preferred by BICEP2.
The potential is approximated in the inflationary region by a general exponentially-corrected plateau, which can be obtained from a generalized 
form of the Starobinsky model, with an infinite series of $R^p$ corrections that sum to a function $f(R)$. 

The functions $f(x)$ and $g(x)$ were analyzed from the point of view of needing to interpolate between the inflationary region and the Higgs potential, 
and in the inflationary region should arise in a consistent quantum gravity theory. 
Of course, specific functions would arise in the case of specific models, that would describe in particular how does the $SO(1,1)$ invariance 
get broken at low energies, but we did not attempt here to construct such models. That is left for future work.
If we modify the non-minimal coupling $\xi$ of the scalar to gravity away from the conformal point, the Starobinsky line is a strong attractor, 
as in the original Starobinsky point, so it seems that the local Weyl invariance is not really essential to the inflationary predictions. 

{\bf Note}. While this paper was being written, the paper \cite{Hertzberg:2014aha} appeared, which also discusses the 
possibility of deforming the $SO(1,1)$ symmetry in a more general way. 
We would like to thank the referee for pointing out to us references \cite{Ellis:2013nxa}, where the general potential (\ref{generalpot}) was 
analyzed, and \cite{Ferrara:2013rsa,Kallosh:2013yoa,Cecotti:2014ipa}, where it was obtained in the context of supergravity models. We have been 
instead focusing on getting this potential from the conformal inflation from the Higgs. 

\section*{Acknowledgements}

We would like to thank Rogerio Rosenfeld for many enlightning discussions and useful comments on the manuscript. 
We would also like to thank Eduardo Pont\'{o}n for discussions. 
The work of HN is supported in part by CNPq grant 301219/2010-9 and FAPESP grant 2013/14152-7, and RC is supported by CAPES.

\bibliography{Higgsinflation}
\bibliographystyle{utphys}

\end{document}